\documentclass[aps,pra,twocolumn,amsmath,amssymb,superscriptaddress,showpacs]{revtex4-1}
\usepackage{epsfig}
\usepackage{graphicx}
\usepackage{color}

\begin{document}

\title{Quantum phase transitions in a generalized compass chain
with three-site interactions}

\author{Wen-Long You}
\affiliation{College of Physics, Optoelectronics and Energy, Soochow
University, Suzhou, Jiangsu 215006, People's Republic of China}

\author{Yu-Cheng Qiu}
\affiliation{College of Physics, Optoelectronics and Energy, Soochow
University, Suzhou, Jiangsu 215006, People's Republic of China}

\author{Andrzej M. Ole\'s}
\affiliation{Max-Planck-Institut f\"ur Festk\"orperforschung,
             Heisenbergstrasse 1, D-70569 Stuttgart, Germany }
\affiliation{Marian Smoluchowski Institute of Physics, Jagiellonian University,
             prof. S. \L{}ojasiewicza 11, PL-30348 Krak\'ow, Poland }

\date{15 March 2016}

\begin{abstract}
We consider a class of one-dimensional compass models with
XYZ$-$YZX-type of three-site exchange interaction in an external
magnetic field. We present the exact solution derived by means of
Jordan-Wigner transformation, and study the excitation gap, spin
correlations, and establish the phase diagram. Besides the
canted antiferromagnetic and polarized phases, the three-site
interactions induce two distinct chiral phases, corresponding to
gapless spinless-fermion systems having two or four Fermi points.
We find that the $z$ component of scalar chirality operator can act
as an order parameter for these chiral phases. We also find that the
thermodynamic quantities including the Wilson ratio can characterize
the liquid phases. Finally, a nontrivial magnetoelectric effect is
explored, and we show that the polarization can be manipulated by the
magnetic field in the absence of electric field.\\
\textit{ Published in Physical Review B 93, 214417 (2016). }
\end{abstract}

\maketitle

\section{Introduction}

The rapid development of spin-orbital physics and quantum information
in recent years motivates the search for the realizations of
intrinsically frustrated orbital (or pseudospin) interactions. Such
interactions lead to radically different behavior from Heisenberg SU(2)
isotropic exchange, and have been in the focus of very active research
in recent years. It was realized that the quantum nature of orbital
degrees of freedom, that may be released by emerging spin-orbital
coupling and spin-orbital entanglement, is interdisciplinary and plays
a crucial role in the fields of strongly correlated electrons
\cite{Feiner,Tok00,Ole05,Kha05,Ole12,Kim09,Jac09,Cha10,Woh11,Brz15}
and cold atoms \cite{Zhao08,Wu08,Sun12,Gal13,Zhou15}.

The strong frustration of spin-orbital interactions can be best
understood by considering generic orbital models, in which the
bond-directional interactions provide the building blocks. Among them,
the two-dimensional (2D) compass model defined on a square lattice
\cite{Nus15} and the Kitaev model on a honeycomb lattice \cite{Kitaev}
can be treated as two quintessential pseudospin models, where the
effective moments cannot simultaneously align to satisfy interactions
with all neighbors as they favor the quantum states with distinct
quantization axes. In fact, the latter model is a rare example of an
interacting 2D spin model that can be rigorously solved, and was found
to support gapped and gapless quantum spin liquids with emergent
fractional excitations obeying non-Abelian statistics.
Otherwise, exact solutions for 2D models with frustrated exchange
exist only for classical Ising interactions where a phase
transition at finite temperature is found \cite{Lon80}. Recent studies
show that also for the 2D compass model a phase transition to nematic
order occurs at finite but much lower temperature \cite{Wen10}.

In low-dimensional magnetic systems collective quantum phenomena are
particularly strong since the reduced dimensionality amplifies the
consequences of frustrated interactions between individual spins. To
probe the exotic phases resulting from bond-directional interactions,
we introduced a one-dimensional (1D) generalized compass model (GCM)
with antiferromagnetic exchange alternating between even and odd bonds
\cite{You1}. Such a model may be realized in layered structures of
transition metal oxides, with alternating exchange interactions along
the bonds parallel to $a$ and $b$ axes along a zigzag chain in an
$(a,b)$ plane \cite{Xiao}, optical lattices \cite{Simon,Str11},
trapped ions \cite{Por04,Kim10},
and coupled photonic cavities \cite{Har08,Chen10}.

On the other hand, the community focuses on two-body interactions in
most systems studied, as they contribute to superexchange and are
readily accessible experimentally. However, the range of the
hybridization of the electron wave function will be finite in some
realistic bonding geometries, and the effect of such long-range
interactions must be addressed.

Recently three-site interactions received considerable attention in
a bit diverse context
\cite{Got99,Tit03,Lou04,Kro08,Cheng10,Der11,Li11,Liu12,Top12,Zhang15,Lei15,Men15,Ste15,Lah15}.
They also occur in an effective spin model in a magnetic field obtained
from a 1D plaquette orbital model by an exact transformation, with spin
dimers that replace plaquettes. Indeed, they are coupled along the
chain by three-spin interactions in the Hilbert space reduced by a
factor of 2 per plaquette \cite{Brz14}. Such complex interactions
between three subsequent sites essentially enrich the ground state
phase diagram of the spin model and open new opportunities for
underlying physics. Experimentally, it can be realized in NMR quantum
simulators \cite{Tseng99,Peng09} or optical lattices \cite{Pachos04}.
Three-site spin interactions have been exhibited the multiferroics
\cite{Suzuki71} and the magnetoelectric effect \cite{Top12,Men15}.

The purpose of this paper is to focus on a 1D GCM with three-site
interactions. We show that this model is exactly solvable and explore
the consequences of three-site interactions. By investigating spin
correlations we identify two chiral phases and demonstrate the
existence of a nontrivial magnetoelectric effect.

The organization of the paper is as follows. In Sec. \ref{sec:ham} we
introduce the Hamiltonian of the 1D GCM with three-site interactions
in Sec. \ref{model} and then present the procedure to solve it exactly
by employing Jordan-Wigner transformation in Sec. \ref{exact}.
The ground state and energy gap are retrieved. In Sec. \ref{sec:cor}
we use spin correlations to characterize each phase and quantum phase
transitions (QPTs). The model in the magnetic field is analyzed in Sec.
\ref{sec:field}, and the complete phase diagram is obtained when the
three-site interactions and magnetic fields are varied. The obtained
exact solution allows us to present the thermodynamic properties
including the Wilson ratio in Sec. \ref{sec:T}.
We also point out that the three-site interactions play
a role in the magnetoelectric effect in Sec. \ref{sec:mee}.
A final discussion and conclusions are given in Sec. \ref{sec:summa}.

\section{Generalized 1D compass model}
\label{sec:ham}

\subsection{The model with three-site exchange}
\label{model}

We consider below a 1D chain of $N$ sites with periodic boundary
conditions, with GCM interactions given by
\begin{equation}
H_{\rm GCM}(\theta)= \sum_{i=1}^{N'}
J_{o}\tilde{\sigma}_{2i-1}(\theta)\tilde{\sigma}_{2i}(\theta)
+J_{e}\tilde{\sigma}_{2i}(-\theta)\tilde{\sigma}_{2i+1}(-\theta).
\nonumber \\
\label{Hamiltonian1}
\end{equation}
Here $N'=N/2$ is the number of two-site unit cells, while $J_o$
and $J_e$ denote the coupling strengths on odd and even bonds,
respectively (below we take $J_o$ as the unit of exchange interaction).
The operator $\tilde{\sigma}_i(\theta)$ (with a tilde) is defined as
a linear combination of $\{\sigma_{i}^x,\sigma_{i}^y\}$ pseudospin
components (Pauli matrices),
\begin{eqnarray}
\tilde{\sigma}_{i}(\theta)&\equiv& \cos(\theta/2)\,\sigma_{i}^x
+\sin(\theta/2)\,\sigma_{i}^y.
\label{tilde}
\end{eqnarray}
These linear combinations imply that Ising-like interactions on an
odd/even bond in Eq. (\ref{Hamiltonian1}) are characterized by the
preferential easy axes selected by an arbitrary angle $\pm\theta/2$.
With increasing angle $\theta$, frustration gradually increases when
the model Eq. (\ref{Hamiltonian1}) interpolates between the Ising model
at $\theta=0$ and the quantum compass model (QCM) at $\theta=\pi/2$, in
analogy to the 2D compass model \cite{Cin10}. The model was solved
exactly and the ground state is found to have order along the easy axis
as long as $\theta\neq \pi/2$, whereas it becomes a highly disordered
spin-liquid ground state at $\theta=\pi/2$ \cite{Brz07,You2}. Here we
introduce the XZY$-$YZX type of three-site interactions in addition,
\begin{equation}
H_{\rm 3-site} =J^*\sum_{i=1}^{N}
(\sigma^x_{i-1}\sigma^z_{i}\sigma^y_{i+1}
-\sigma^y_{i-1}\sigma^z_{i}\sigma^x_{i+1}),
\label{Hamiltonian2}
\end{equation}
where $J^*$ is its strength. Such interactions between three adjacent
sites emerge as an energy current of a compass chain in the
nonequilibrium steady states, as discussed in the Appendix.

The complete Hamiltonian of the 1D GCM with the three-site XZY$-$YZX
interaction is
\begin{eqnarray}
{\cal H} =H_{\rm GCM}+H_{\rm 3-site}.
\label{model}
\end{eqnarray}

\subsection{Exact solution}
\label{exact}

We employ the Jordan-Wigner transformation which maps explicitly
between quasispin operators and spinless fermion operators through
the following relations \cite{Bar70}:
\begin{eqnarray}
\sigma _{j}^{z}& =&1-2c_{j}^{\dagger }c_{j}, \quad
\sigma _{j}^{y}=i\sigma _{j}^{x}\sigma _{j}^{z}, \notag \\
\sigma _{j}^{x}& =& \prod_{i<j}\,(1-2c_{i}^{\dagger }c_{i}^{})
 (c_{j}^{}+c_{j}^{\dagger}),
 \label{JW}
\end{eqnarray}
where $c_{j}$ and $c_{j}^{\dagger }$ are annihilation and creation
operators of spinless fermions at site $j$ which obey the standard
anticommutation relations, $\{c_{i},c_{j}\}=0$ and
$\{c_{i}^{\dagger},c_{j}\}=\delta_{ij}$. By substituting Eqs.
(\ref{JW}) into Eq. (\ref{model}), we arrive at a simple bilinear
form of the Hamiltonian \eqref{model} in terms of spinless fermions:
\begin{eqnarray}
\cal{H}&=&
\sum_{i=1}^{N'} \Big[J_{o}  e^{i\theta}
    c_{2i-1}^{\dagger} c_{2i}^{\dagger}
+  J_{o}  c_{2i-1}^{\dagger} c_{2i}^{} \nonumber \\
& &\hskip .5cm + J_{e}e^{-i\theta} c_{2i}^{\dagger} c_{2i+1}^{\dagger}
+  J_{e}  c_{2i}^{\dagger} c_{2i+1}^{} \nonumber \\
& &\hskip .5cm - 2iJ^*(c_{2i-1}^{\dagger} c_{2i+1}+
c_{2i}^{\dagger}c_{2i+2}^{})+{\rm H.c.}\Big].
\end{eqnarray}
Next discrete Fourier transformation for plural spin sites is
introduced by
\begin{eqnarray}
c_{2j-1}\!=\frac{1}{\sqrt{N'}}\sum_{k}e^{-ik j}a_{k},\text{ \ \ }
c_{2j}\!=\frac{1}{\sqrt{N'}}\sum_{k}e^{-ik j}b_{k},
\end{eqnarray}
with discrete momenta as
\begin{eqnarray}
k=\frac{n\pi}{ N^\prime}, \quad
n= -(N^\prime\!-1), -(N^\prime\!-3),\ldots, (N^\prime\! -1).
\end{eqnarray}
The Hamiltonian takes the following form, which is suitable
to introduce the Bogoliubov transformation:
\begin{eqnarray}
\cal{H}&=& \sum_{k} \left[ B_k^{} a_{k}^{\dagger}
b_{-k}^{\dagger}+ A_k^{} a_{k}^{\dagger} b_{k}^{}
- A_k^* a_{k}^{}b_{k}^{\dagger}-B_k^* a_{k}^{}b_{-k}^{} \right.
\nonumber \\
& & \left.\hskip .5cm - 4J^* \sin k (a_{k}^{\dagger} a_{k}^{}
+ b_{k}^{\dagger} b_{k}^{})\right].
\label{Hamiltonian5}
\end{eqnarray}
where
\begin{eqnarray}
A_k&=&  J_{o} +  J_{e}+   e^{ik}, \nonumber\\
B_k&=& J_o e^{i\theta}-J_e e^{i(k-\theta)}.
\end{eqnarray}
To diagonalize the Hamiltonian Eq. (\ref{Hamiltonian5}), we rewrite it
in the Bogoliubov-de Gennes form,
\begin{eqnarray}
{\cal H} &=&  \sum_{k}\,
\Gamma_k^{\dagger}\,\hat{M}_k^{}\,\Gamma_k^{},
\label{FT2}
\end{eqnarray}
where
\begin{eqnarray}
\hat{M}_k\!=\frac{1}{2}\!\left(\!
\begin{array}{cccc}
-G_k & 0 &   S_k   &  P_k+Q_k  \\
0 & -G_k & P_k- Q_k   &    -S_k  \\
 S_k^*  &  P_k^*-Q_k^*   & -G_k & 0 \\
P_k^*+Q_k^*   &  -S_k^*   & 0 &-G_k
\end{array}\!\right),
\label{MainHamMatrix}
\end{eqnarray}
and $\Gamma_k^{\dagger}=(a_k^{\dagger},a_{-k}^{},b_k^{\dagger},b_{-k}^{})$.
In Eq. (\ref{MainHamMatrix}) the compact notation is introduced:
\begin{eqnarray}
P_k&=&-i (J_e e^{ik}+J_o)\sin\theta, \nonumber \\
Q_k&=& (J_e e^{ik}-J_o)\cos\theta,   \nonumber \\
S_k &=&J_o+J_e e^{ik}, \nonumber \\
G_k&=& 2J^* \sin k.
\label{compactnotations}
\end{eqnarray}

\begin{figure}[t!]
\includegraphics[width=\columnwidth]{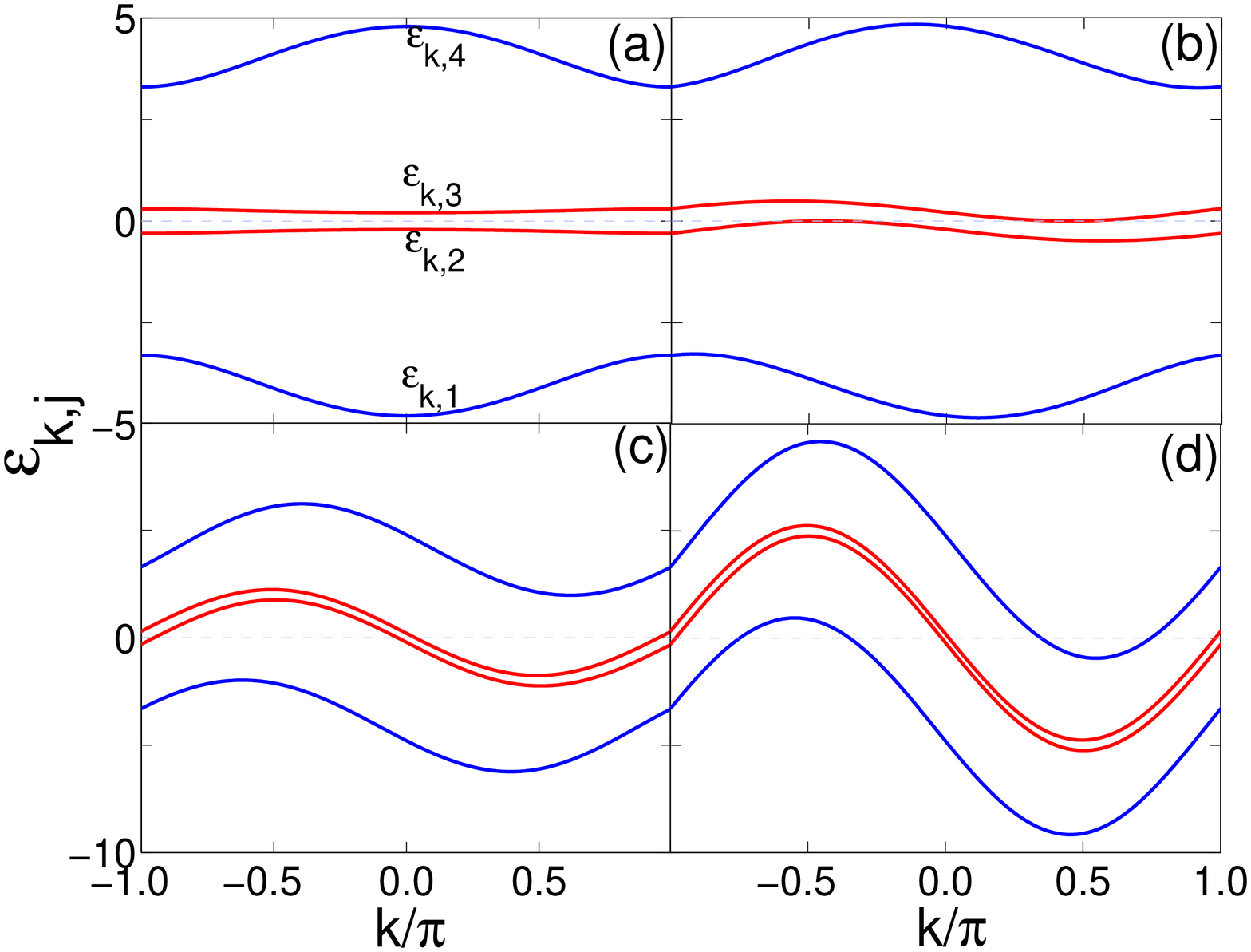}
\caption{
The energy spectra $\varepsilon_{k,j}$ ($j=1,\cdots,4$) for
increasing $J^*$:
(a) $J^*$ = 0,
(b) $J^*$ = 0.239,
(c) $J^*$ = 2, and
(d) $J^*$ = 5.
Parameters are as follows: $J_o = 1$, $J_e = 4$, $\theta=\pi/3$. }
 \label{Fig1:spec}
\end{figure}

The diagonalization of Hamiltonian (\ref{MainHamMatrix}) is achieved by
a four-dimensional Bogoliubov transformation which connects the operators
$\{a_k^{\dagger},a_{-k}^{},b_k^{\dagger},b_{-k}^{}\}$ with four kind of
quasiparticles,
$\{\gamma_{k,1}^{\dagger},\gamma_{k,2}^{\dagger},
   \gamma_{k,3}^{\dagger},\gamma_{k,4}^{\dagger}\}$,
\begin{eqnarray}
\left(
\begin{array}{c}
\gamma_{k,1}^{\dagger} \\
\gamma_{k,2}^{\dagger}  \\
\gamma_{k,3}^{ \dagger}\\
\gamma_{k,4}^{\dagger}
\end{array}
\right)=\hat{U}_{k} \left(
\begin{array}{c}
a_k^{\dagger}  \\
a_{-k}   \\
b_k^{\dagger}   \\
b_{-k}
\end{array}\right),
\label{eq:2DXXZ_RDM}
\end{eqnarray}
where the rows of $\hat{U}_{k}$ are eigenvectors of the
Bogoliubov-de Gennes equations. The diagonalization is readily
performed to yield the eigenspectra $\varepsilon_{k,j}$
($j=1,\cdots,4$):
\begin{eqnarray}
\varepsilon_{k,1(2)}=-\frac{1}{2}\left(
\sqrt{\xi_k \pm \sqrt{\xi_k^2-\tau_k^2}}+G_k\right), \nonumber \\
\varepsilon_{k,3(4)}=\frac{1}{2}\left(
\sqrt{\xi_k \mp \sqrt{\xi_k^2-\tau_k^2}}-G_k\right),
\label{excitationspectrum}
\end{eqnarray}
where
\begin{eqnarray}
\xi_k&=&\vert P_k \vert^2 + \vert Q_k \vert^2 +  \vert S_k \vert^2 ,
\nonumber \\
\tau_k&=&\vert  P_k^2 - Q_k^2  + S_k^2 \vert.
\end{eqnarray}
The eigenenergies for various $J^*$ are labeled sequentially from the
bottom to the top as $\varepsilon_{k,1},\cdots,\varepsilon_{k,4}$ in
Fig. \ref{Fig1:spec}. One finds that finite $J^*$ removes the symmetry
of the spectra with respect to $\varepsilon=0$ energy and they are
not invariant with respect to the $k\to -k$ transformation, in contrast
to the case of the GCM with $J^*=0$ shown in Fig. \ref{Fig1:spec}(a).
The three-site interactions break both parity (P) symmetry and time
reversal (T) symmetry. Note that modes $k= 0,\pm \pi$ are time reversal
invariant and their excitations are independent of $J^*$ as a
consequence of vanishing $G_{k}$. Instantly, we obtain the diagonal
form of the Hamiltonian,
\begin{eqnarray}
{\cal H}=\sum_{k }\sum_{j=1}^{4}  \varepsilon_{k,j}\,
 \gamma_{k,j}^{\dagger}\gamma_{k,j}^{} .
\label{diagonalform}
\end{eqnarray}

The most important properties of the 1D quantum system can be explored
in the ground state. The ground state of any fermion system follows the
Fermi-Dirac statistics, and the lowest energy is obtained when all the
quasiparticle states with negative energies are filled by fermions.
More precisely, in the thermodynamic limit ($N\to\infty$) the ground
state of the system, $|\Phi_0\rangle$, corresponds to the configuration
with chemical potential $\mu=0$, where all the states with
$\varepsilon_{k,j}<0$ are occupied and the ones with
$\varepsilon_{k,j}\ge 0$ are empty. This state is realized by means of
the corresponding occupation numbers,
\begin{equation}
n_{k,j}=\langle\Phi_0\vert\gamma_{k,j}^{\dagger}\gamma_{k,j}^{}
\vert\Phi_0\rangle = \left\{
  \begin{array}{l l}
    0 & \quad {\rm for}\;\varepsilon_{k,j} \ge 0,\\
    1 & \quad{\rm for}\;\varepsilon_{k,j}<0.
  \end{array} \right.
\end{equation}
One recognizes that the Bogoliubov-de Gennes Hamiltonian
(\ref{MainHamMatrix}) actually acts in an artificially enlarged
Nambu-spinor space and it respects an emergent particle-hole symmetry
(PHS) ${\cal C}$, i.e., ${\cal C}\hat{M}_k{\cal C}=\hat{M}_{-k}$, with
${\cal C}^2=1$. Here in the so-called particle-hole space, the extra
degree of freedom $\textbf{C}^{2}$ leads to two copies of the actual
excitation spectrum, a particle and a hole copy emerge simultaneously.
The PHS implies here that $\gamma_{k,4}^{\dagger}$=$\gamma_{-k,1}^{}$,
$\gamma_{k,3}^{\dagger}$=$\gamma_{-k,2}^{}$, as is evidenced in Fig.
\ref{Fig1:spec}. The bands with positive energies correspond to the
electron excitations while the negative ones are the corresponding
hole excitations. When all quasiparticles above the Fermi surface are
absent the ground state energy may be expressed as:
 \begin{eqnarray}
E_0 = -\frac{1}{2} \sum_{k} \sum_{j=1}^4
\left\vert \varepsilon_{k,j} \right\vert.
\label{E0expression}
\end{eqnarray}
Accordingly, the gap is determined by the absolute value of the
difference between the second and third energy branches,
\begin{equation}
\Delta=\min_{k}
\left\vert\varepsilon_{k,2}-\varepsilon_{-k,3}\right\vert.
\end{equation}

One finds that with the increase of $J^*$, the minimum of
$\varepsilon_{k,3}$ bends down until it touches $\varepsilon=0$ when
$J^*$ reaches a threshold value $J_{c,1}^*$, i.e., $\Delta$ = 0; cf.
Fig. \ref{Fig1:spec}(b).
A gapless mode shows up at some incommensurate mode $k_{ic}$ and the
spectrum vanishes quadratically. Further increase of $J^*$ leads to
the bands inversion between portions of $\varepsilon_{k,2}$ and
$\varepsilon_{k,3}$. There is a negative-energy region of
$\varepsilon_{k,3}$ in $k$ space shown in Fig. \ref{Fig1:spec}(c),
and there are two Fermi points across the Fermi surface. When $J^*$
exceeds another threshold value $J_{c,2}^*$ the energy spectrum of
spinless fermions may also have two additional Fermi points
\cite{Lou04}, as observed in Fig. \ref{Fig1:spec}(d).
A Lifshitz transition occurs following the topological
change of the Fermi surface in the Brillouin zone.

\section{Correlations and quantum phase transitions}
\label{sec:cor}

\begin{figure}[t!]
\begin{center}
\includegraphics[width=\columnwidth]{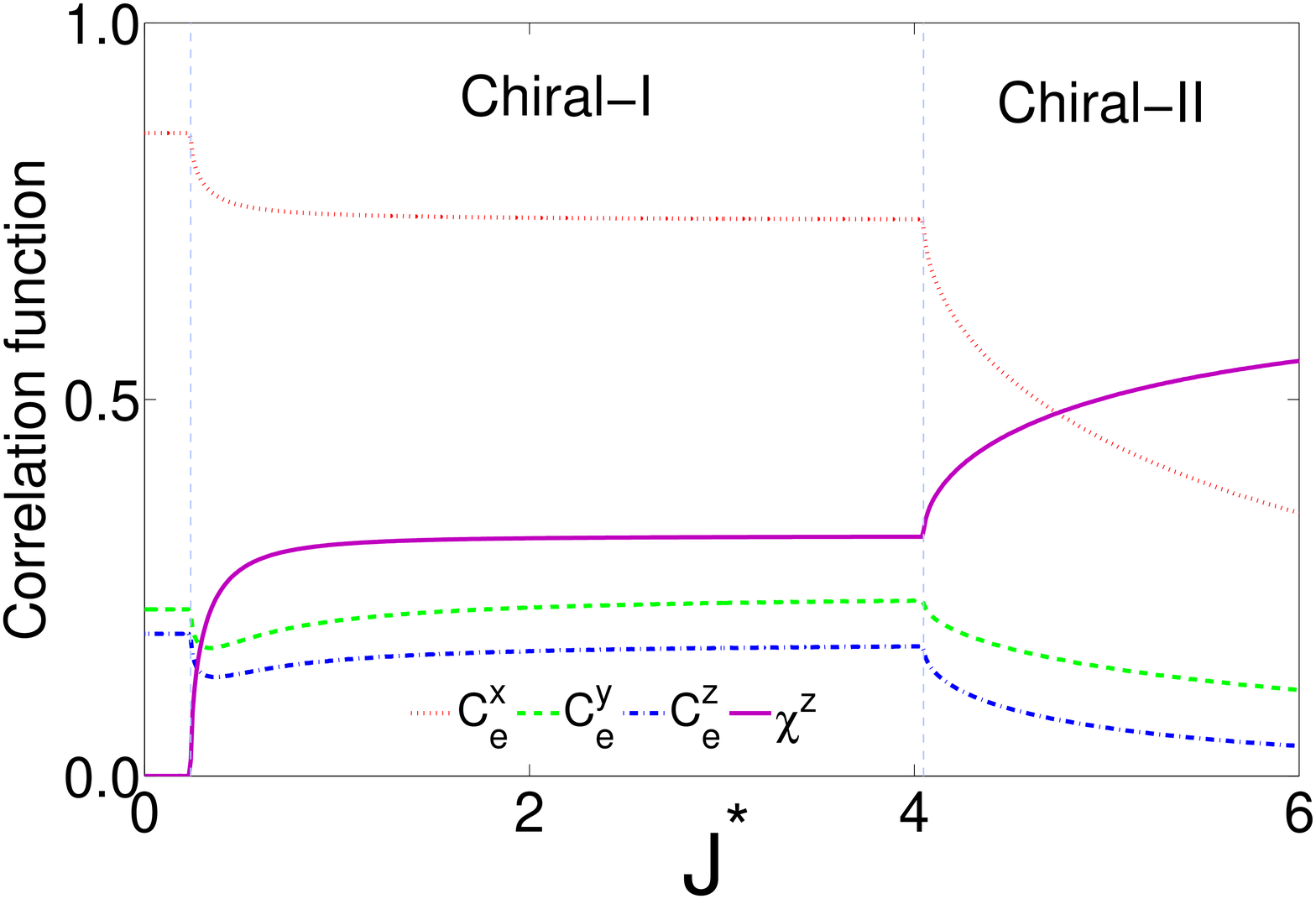}
\end{center}
\caption{  The nearest neighbor correlations $C^\alpha$
on even bonds and chirality $\chi^\alpha$ by increasing $J^*$ for $h=0$.
Parameters are as follows: $J_o = 1$, $J_e = 4$, $\theta=\pi/3$. }
\label{Fig:CF}
\end{figure}

In order to characterize the QPTs, we studied the nearest neighbor
spin correlation function defined by
\begin{eqnarray}
C^{\alpha}_{l}&=&-\frac{2}{N}\sum_{i=1}^{N'}\langle
\sigma_{i}^\alpha
\sigma_{i+l}^\alpha \rangle,
\end{eqnarray}
where $l$=1(-1) and
the superscript $\alpha=x,y,z$ denotes the cartesian component,
and the $z$ component of scalar chirality operator \cite{Wen89}
\begin{eqnarray}
{\cal \chi}^{z} = -\frac{1}{N}\sum_{i=1}^{N} \langle {\sigma}_{i}^z
\vec{z}\cdot [\vec{\sigma}_{i-1}\times\vec{\sigma}_{i+1}].\rangle
\label{chir}
\end{eqnarray}
The scalar chirality operator can act as a local order parameter for
states without PT symmetry. As shown in Fig. \ref{Fig:CF}, the
ground state has finite nearest neighbor correlation functions for
$J^*=0$, among which $x$ components $\{C_l^x\}$ dominate for
$\theta=\pi/3$, implying that the adjacent spins are antiparallel
and aligned with a canted angle with respect to the $x$ axis.
Indeed, the ground state of the GCM is a canted N\'eel (CN) phase for
$\theta<\pi/2$.

With the increase of $J^*$, the nearest neighbor correlation functions
remain constant. After $J^*$ surpasses $J_{c,1}^*$, the system stays
in a chiral-I phase without finite energy gap, characterized by a
nonzero ${\cal \chi}^{z}$. In such a chiral-I phase, $x$ components
$C_l^x$ decrease while $C_l^y$ and $C_l^z $ grow as $J^*$ increases,
but they become saturated quickly. When $J^*>J_{c,2}^*$, the system
enters chiral-II phase, where ${\cal \chi}^{z}$ grows rapidly and
$\{C_l^\alpha\}$ ($\alpha=x$,$y$, and $z$) decreases simultaneously.

In the fermionic picture different phases correspond to different
Fermi-surface topology (different number of Fermi points) for fermions.
In particular, the two Fermi-point spinless fermions (chiral-I phase)
is distinct from the four-Fermi-point spinless fermions (chiral-II
phase) \cite{Lou04}. Both spin-liquid phases have gapless excitations,
however, the appearance of new points $k_F$ in the Fermi surface when
the controlling parameter crosses a critical value will witness
a general feature of the discontinuities in the correlation functions.
We remark that the number of gapless modes determine the effective
central charge and the coefficients of the area-law violating term of
bipartite entanglement entropy \cite{Eisert,Eloy12}. Notably, the
chiral-II phase is a dedicated phase of the critical XX model with
three-site XZY$-$YZX interactions added
\cite{Lou04,Liu12,Der11,Kro08,Tit03,Top12}, while this phase is absent
for anisotropic XY model \cite{Lei15}. Here we observe the three-site
XZY$-$YZX interactions in the GCM surprisingly triggers both chiral
states for arbitrary $\theta$, and two different Tomonaga-Luttinger
liquids reflect the importance of Fermi surface topology.

\begin{figure}[b!]
\includegraphics[width=7cm]{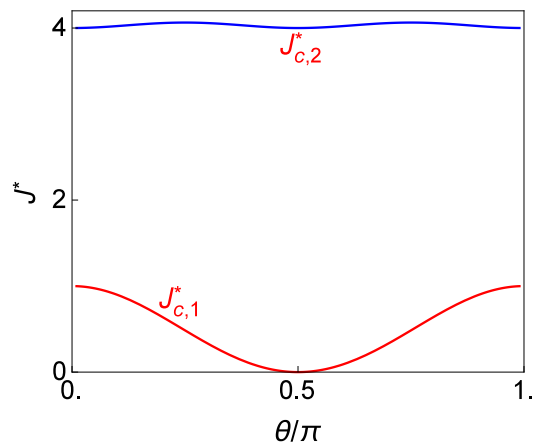}
\caption{
The critical value of $J^*$ as a function of $\theta$.
Parameters are as follows:  $J_o$=1, $J_e$=4. }
\label{Lc}
\end{figure}

The determination of critical values of $J_{c,1}^*,J_{c,2}^*$ and the
corresponding incommensurate momentum $k_{ic}$ can be given by
\begin{eqnarray}
\varepsilon_{k_{ic},3(4)}=0 ,
\quad   \partial \varepsilon_{k_{ic},3(4)}/\partial k=0.
\label{eqcrit}
\end{eqnarray}
This leads to the following quartic equation for $x_{ic}=\cos k_{ic}$:
\begin{eqnarray}
 x_{ic}^4   + c_3 x_{ic}^3   + c_2 x_{ic}^2 +c_0  =0,
\end{eqnarray}
where
\begin{eqnarray}
c_3&=&4(J_o^2 + J_e^2)/(3J_o J_e \sin^2\theta),   \nonumber \\
c_2&=&(J_o^2+J_e^2)^2/(3J_o^2 J_e^2 \sin^4\theta)-4\cot^4\theta/3+2/3,
\nonumber \\
c_0&=&-1/3. \nonumber
\end{eqnarray}
This quartic equation can be solved analytically but the form is rather
contrived. We plot the critical lines with respect to $\theta$ in Fig.
\ref{Lc}. One finds that in the Ising limit, i.e., for $\theta \to 0$,
it yields
\begin{eqnarray}
J^*_{c,1} \to \textrm{min} (J_o, J_e) \quad {\rm and} \quad
J^*_{c,2} \to \textrm{max} (J_o, J_e).
\end{eqnarray}
While in the compass limit, i.e., for $\theta \to \pi/2$, we have
\begin{eqnarray}
J^*_{c,1} \to  0 \quad {\rm and} \quad
J^*_{c,2} \to \textrm{max} (J_o, J_e).
\end{eqnarray}
In other words, the system for $\theta=\pi/2$ has an emergent
$\mathbb{Z}_2$ symmetry and the ground state can not be ordered.
Any infinitesimal perturbation of $J^*$ will induce the system into
gapless chiral-I state. For the parameters we choose mostly in this
paper, i.e., $J_o=1$, $J_e=4$, $\theta=\pi/3$, one finds
$J^*_{c,1}=0.239$ and $J^*_{c,2}=4.048$.

\section{Effect of transverse field}
\label{sec:field}

We now consider the case where the magnetic field $h$ is perpendicular
to the easy plane of the spins, i.e., $\vec{h}=h\hat{z}$.
In this case, the Zeeman term is given by
\begin{eqnarray}
{\cal H}_h=
h\hat{z}\cdot\sum_{i=1}^{N'}(\vec{\sigma}_{2i-1}+\vec{\sigma}_{2i}),
\end{eqnarray}
where $h$ is the magnitude of the transverse external field.
Subsequently, in Nambu representation, the Hamiltonian matrix
$\hat{M}_k$ (\ref{MainHamMatrix}) is modified in the following way:
\begin{eqnarray}
\hat{M}_k \to \hat{M}_k^{'}=\hat{M}_k -h \mathbb{I}_2 \otimes \sigma^z,
\label{Mk_h}
\end{eqnarray}
where $\mathbb{I}_2$ is a ($2\times 2$) unity matrix. It is obvious
that the external magnetic field plays the role of a chemical potential
for spinless fermions.

\begin{figure}[t!]
\includegraphics[width=\columnwidth]{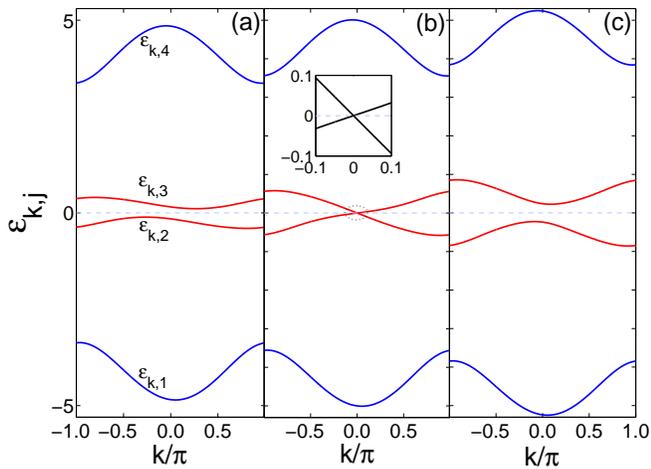}
\caption{
The energy spectra $\varepsilon_{k,j}$ ($j=1,\cdots,4$) for increasing
electric field $h$:
(a) $h$ = 1,
(b) $h$ = 2, and
(c) $h$ = 3.
The inset in (b) is an amplification of the level crossing at the Fermi
energy marked by dashed circle below. Parameters are as follows:
$J_o$=1, $J_e$=4, $\theta=\pi/3$, and $J^*=0.1$. }
\label{Fig2:spec}
\end{figure}

After diagonalization four branches of energies $\varepsilon_{k,j}$,
with $j=1,\cdots,4$, are given by the following expressions:
\begin{eqnarray}
\varepsilon_{k,1(2)}=
-\frac{1}{2}\left(\sqrt{\zeta_k \pm \sqrt{\eta_k }}-G_k\right),
\nonumber \\
\varepsilon_{k,3(4)}=
 \frac{1}{2}\left(\sqrt{\zeta_k \mp \sqrt{\eta_k }}-G_k\right),
\label{excitationspectrum2}
\end{eqnarray}
where
\begin{eqnarray}
\zeta_k&=&\vert P_k \vert^2 + \vert Q_k \vert^2 + \vert S_k
\vert^2+   h^2, \nonumber \\
\eta_k&=& (S_k^* Q_k + S_k Q_k^*)^2
-(S_k^* P_k - S_k P_k^*)^2 \nonumber \\
&+&(P_k^* Q_k + P_k Q_k^*)^2 + 4 \vert S_k \vert^2 h^2.
\end{eqnarray}
The magnetic field further breaks the T symmetry and polarizes spins
along $z$ direction. The analytical solution for $J^*$ = 0 had been
scrutinized recently. One finds that increasing transverse field
induces finite transverse polarization $\langle \sigma_i^z\rangle$ and
drives the system into a saturated polarized phase above the critical
field \cite{You1}. The field-induced QPT is of second order for
arbitrary angle $\theta$ and occurs at at the critical value,
\begin{eqnarray}
h_c = 2 \sqrt{J_o J_e}\cos\theta.
\label{hc}
\end{eqnarray}

\begin{figure}[t!]
\begin{center}
\includegraphics[width=\columnwidth]{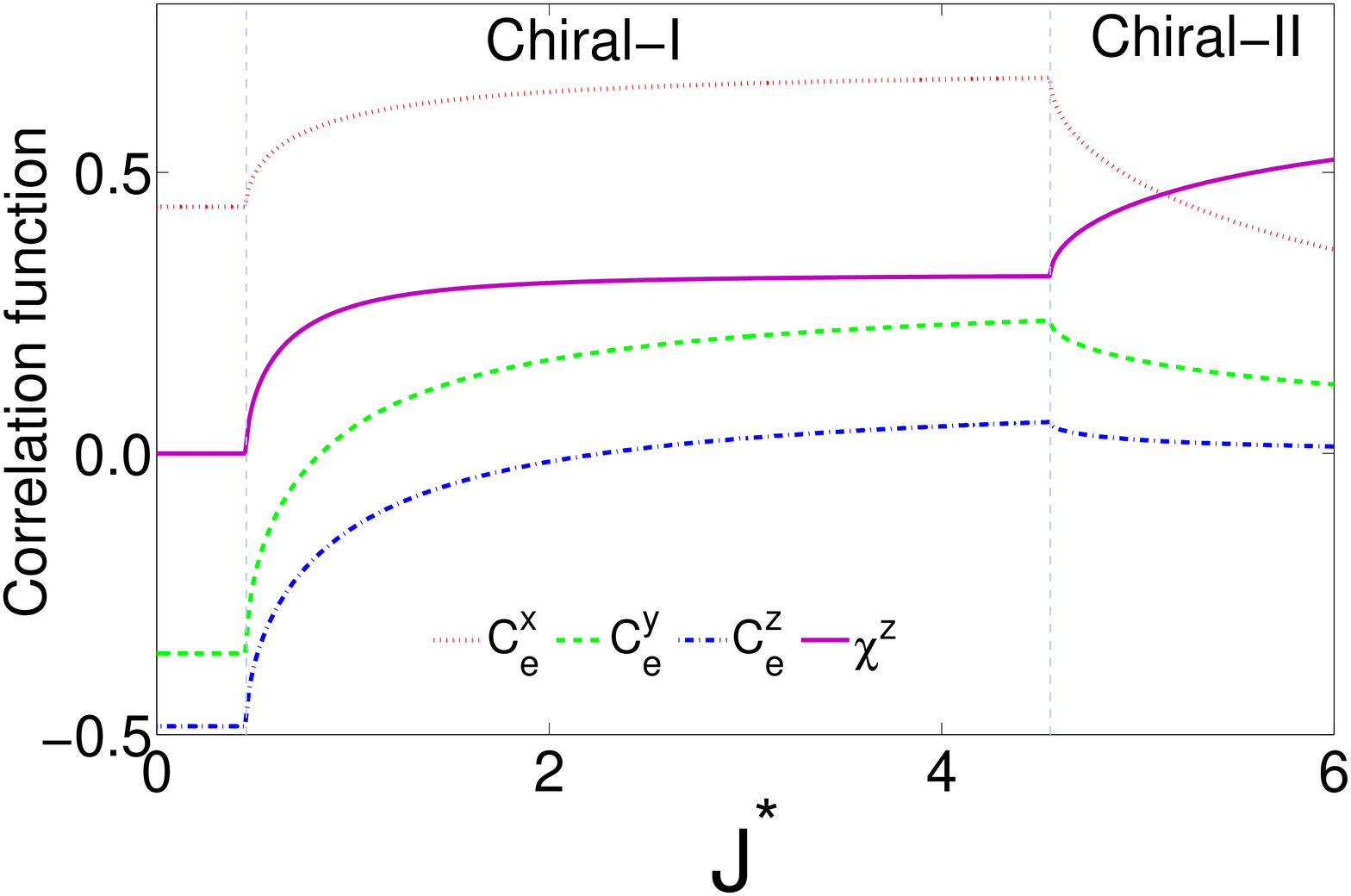}
\end{center}
\caption{  The nearest neighbor correlations $C^\alpha$
on even bonds and chirality $\chi^\alpha$ by increasing $J^*$ for $h = 3$.
Parameters are as follows: $J_o = 1$, $J_e = 4$, $\theta=\pi/3$. }
\label{Fig:CF2}
\end{figure}

The field-induced criticality is suited at momentum $k= 0$, where
$G_k$ does not play a role, see Eq. (\ref{compactnotations}). Figure
\ref{Fig2:spec} shows the energy spectra obtained for three typical
values of $h$ and fixed weak $J^*=0.1$. We find that a finite gap
separates occupied from empty bands except when $h=h_c$, see
Eq. (\ref{hc}). A small value of $J^*$ does not modify the critical
field and the gap vanishes linearly for $\theta\neq\pi/2$, see inset
in Fig. \ref{Fig2:spec}(b). When $h=h_c$ the gaps opens and grows with
increasing $(h-h_c)$, see Fig. \ref{Fig2:spec}(c).

\begin{figure}[t!]
\includegraphics[width=\columnwidth]{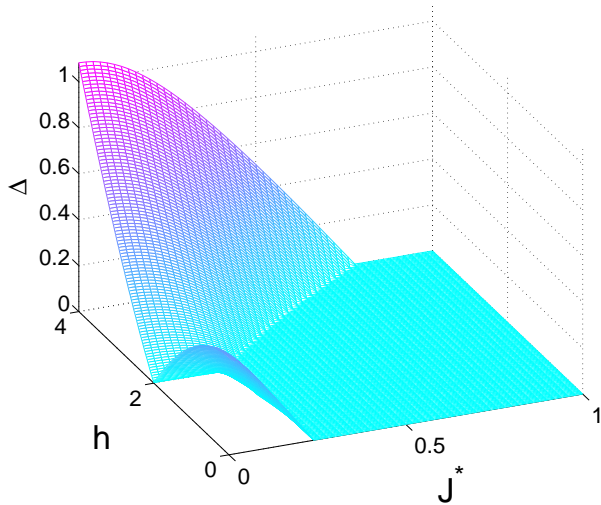}
\caption{  The gap $\Delta$ as a function of $h$ and $J^*$.
Parameters are as follows: $J_o$=1, $J_e$=4, $\theta=\pi/3$. }
\label{Fig1:gap}
\end{figure}

The nearest neighbor correlation functions $\{C_l^\alpha\}$
($\alpha$=$x$, $y$, and $z$) and the $z$ component of scalar chirality
operator ${\cal \chi}^z$ for increasing $J^*$ at $h=3$ are shown in
Fig. \ref{Fig:CF2}. Finite magnetic field expands the range of CN phase
and increases both $J_{c,1}^*$ and $J_{c,2}^*$, see Fig. \ref{Fig:CF2}.
The $z$ components $\{C_l^z\}$ dominate over $x$ components $\{C_l^x\}$
for small $J^*$ and $\theta=\pi/3$, suggesting that the spins are
aligned along the $z$ axis according to the sign of $\{C_l^z\}$.
The correlation functions are found to be almost independent of $J^*$
as long as the system is within the polarized state, but they change
in a discontinuous way at phase transitions. As $J^*$ rises above the
critical value $J_{c,1}^*$, a nonzero chirality ${\cal \chi}^{z}$
starts to grow and saturates. One finds that $C_l^y$ and $C_l^z$
decrease and change sign from negative to positive values upon
increasing $J^*$, which is contrast to the trend observed for $C_l^x$.
A sharp upturn of ${\cal \chi}^{z}$ occurs for $J>J_{c,2}^*$,
and it continues to increase with $J^*$. Simultaneously, all the
correlation functions $\{C_l^\alpha\}$ ($\alpha=x,y,z$) decrease
strongly towards zero when the system enters the chiral-II phase.

To present a three-dimensional panorama of the excitation gap,
we display $\Delta$ for varying $h$ and $J^*$ in Fig. \ref{Fig1:gap}.
The gap $\Delta$ diminishes for large value of $J^*$.

\begin{figure}[t!]
\includegraphics[width=\columnwidth]{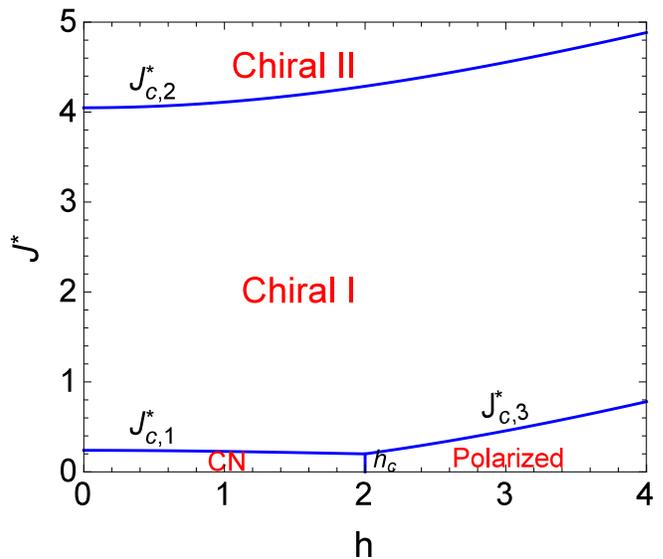}
\caption{
Magnetic phase diagram of the 1D GCM as a function of transverse field
$h$ and three-site XZY$-$YXZ interaction $J^*$.
Parameters are as follows: $J_o$=1, $J_e$=4, $\theta=\pi/3$. }
\label{phd2}
\end{figure}

Similarly, we can discriminate the critical lines $J^*_{c,1(2)}$ and
zero-gap modes $k_{ic}$ using the relations in Eq. (\ref{eqcrit}).
The phase diagram is shown in Fig. \ref{phd2}. The phase diagram at
finite three-site XZY$-$YZX interaction and magnetic field consists
of four phases:
(i) canted antiferromagnetic,
(ii) polarized,
(iii) chiral-I, and
(iv) chiral-II.
A tricritical point is determined by the intersection of both
critical lines which can be obtained analytically:
\begin{eqnarray}
h_c=2\sqrt{J_o J_e}\cos \theta, J_c^*= J_o J_e \cos^2\theta/(J_o+J_e).
\end{eqnarray}
In the special case of $\theta=\pi/2$, the CN phase is never stable.

\section{THERMODYNAMIC PROPERTIES}
\label{sec:T}

Since the exact solution of the GCM with three-site interaction and the
external field is at hand, it is straightforward to obtain its complete
thermodynamic properties at finite temperature. All quantum phase
transitions of the present 1D GCM are of second order. Among many
thermodynamic quantities, the specific heat and magnetic susceptibility
are easy to to be measured, and both of them are proportional to the
electronic density of states at Fermi energy.
For the particle-hole excitation spectrum (\ref{excitationspectrum2}),
the free energy of the quantum spin chain at temperature $T$ reads,
\begin{eqnarray}
{\cal F}= -  k_B T \sum_k\sum_{j=1}^4
\ln\left(2\cosh\frac{\varepsilon_{k,j}}{2k_B T}\right).
\end{eqnarray}
The low temperature behavior of the heat capacity,
\begin{eqnarray}
\label{cv}
C_V(T)&=&-T\left(\frac{\partial^2{\cal F}}{\partial T^2}\right)_h
\nonumber \\
&=& k_B \sum_k \sum_{j=1}^{4} \frac{(\varepsilon_{k,j}/2k_B T)^2}
{  \cosh^2 (\varepsilon_{k,j}/2k_BT)}.
\end{eqnarray}
The magnetic susceptibility is defined as follows,
\begin{eqnarray}
\label{chif}
\chi(T)&=&-\left(\frac{\partial^2{\cal F}}{\partial h^2}\right)_T
- \frac{1}{2 }\sum_k\sum_{j=1}^{4}\left\{
\frac{\partial^2\varepsilon_{k,j}}{\partial h^2}
\tanh\left( \frac{\varepsilon_{k,j}}{2k_BT}\right) \right. \nonumber \\
&+& \left.\left(\frac{\partial\varepsilon_{k,j}}
{\partial h}\right)^2\left[2k_BT\cosh^2\left(\frac{
\varepsilon_{k,j}}{2k_BT}\right)\right]^{-1}\right\}.
\end{eqnarray}

\begin{figure}[t!]
\includegraphics[width=8.4cm]{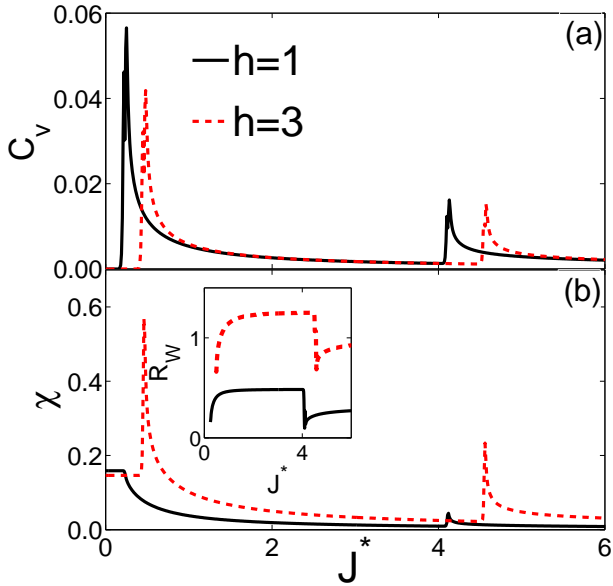}
\caption{
The thermodynamic properties for two values of $h=1$ and $h=3$
at fixed temperature $T=0.01$:
(a) the specific heat $C_V$,
(b) the magnetic susceptibility $\chi$.
The inset shows the Wilson ratio $R_W$ (\ref{wira}) as a function of
three-site XZY$-$YXZ interaction $J^*$ for $h=1$ and $h=3$.
Parameters are as follows: $J_o$=1, $J_e$=4, $\theta=\pi/3$. }
\label{RW}
\end{figure}

At low temperatures the specific heat has a linear dependence on $T$
in liquid metals due to the contribution from the electrons within the
energy interval $k_BT$ near the Fermi surface, while the magnetic
susceptibility is independent of temperature owing to the fact that
only the electrons within the energy $\mu_B gH$ near the Fermi surface
contribute to magnetization.
The Sommerfeld-Wilson ratio (Wilson ratio in short) is a parameter
which characterizes strongly correlated Fermi liquids. It is defined
as a dimensionless ratio of the zero-temperature magnetic
susceptibility $\chi$ and the coefficient of the linear term
$\propto T$ in the electronic specific heat $C_V(T)$ \cite{Wilson},
\begin{eqnarray}
R_{\rm W}=\frac{1}{3}\left(\frac{2\pi k_B}
{\mu_B g_{\rm Lande}} \right)^2\frac{T\chi(T)}{C_V(T)},
\label{wira}
\end{eqnarray}
where $k_B$ is Boltzmann's constant, $\mu_B \equiv e/(2mc) $ is the
Bohr magneton, $g_{\rm Lande}\simeq 2$ is the Lande factor.
Such quantity measures the strength of magnetic
fluctuations versus thermal fluctuations.

Figure \ref{RW} shows the specific heat $C_V(T)$ and the magnetic
susceptibility $\chi(T)$ for increasing $J^*$, in the range which
covers all phases. In a 1D antiferromagnet, the zero-temperature
magnetic susceptibility exhibits a square-root divergence across
critical fields. The Wilson ratio (\ref{wira}) undergoes an increase due
to sudden changes in the density of states near the critical fields
\cite{Gaun13}. $R_{\rm W}=1$ in the free-electron limit when
$J^*\to\infty$. However, we notice that $R_{\rm W}$ deviates from 1 in
chiral-I phase. In particular, $R_{\rm W}$ is larger here than that in
chiral-II phase.
Furthermore, $R_{\rm W}$ is enhanced by increasing magnetic field, see
inset in Fig. \ref{RW}. The Wilson ratio can be measured experimentally
as for instance in a recent experiment on a gapped spin-1/2 Heisenberg
ladder compound (C$_7$H$_{10}$N)$_2$CuBr$_2$~\cite{Nin12}.

\section{Magnetoelectric effect }
\label{sec:mee}

Next we consider the magnetoelectric effect (MEE), where the roles of
magnetization and polarization can be interchanged. A key quantity to
characterize the MEE is the linear magnetoelectric susceptibility which
defines the dependence of magnetization on the electric field, or the
polarization dependence on the magnetic field.

\begin{figure}[t!]
\includegraphics[width=\columnwidth]{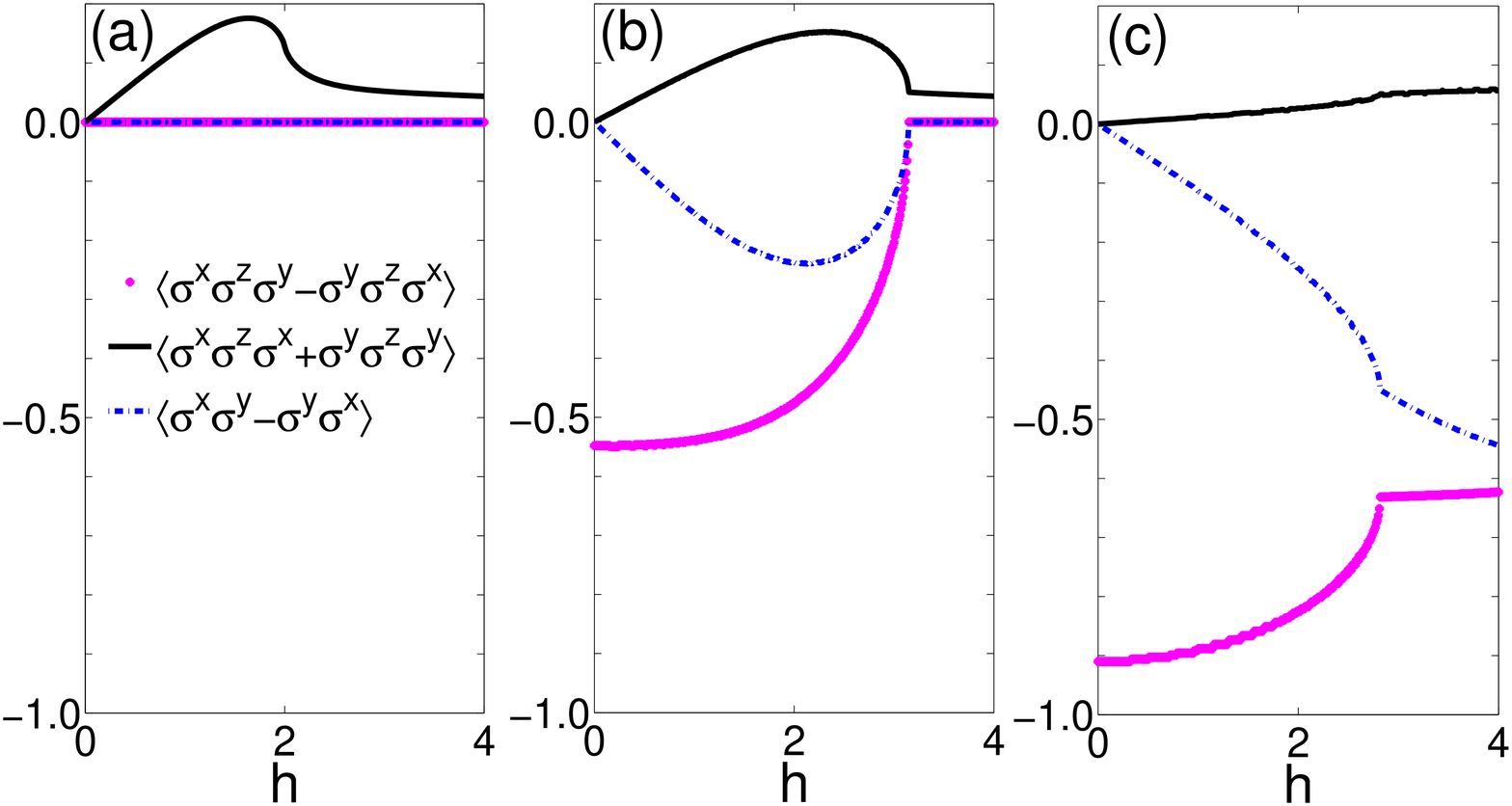}
\caption{
Electric polarizations (see legend) as functions of external field $h$
for:
(a) $J^*=0$,
(b) $J^*=0.5$, and
(c) $J^*=4.5$.
Parameters are as follows: $J_o$=1, $J_e$=4, $\theta=\pi/3$.  }
 \label{OP_h}
\end{figure}

The three-spin interaction was naturally claimed to contribute to the
ferroelectricity in the Katsura-Nagaosa-Balatsky (KNB) formula for its
particular form \cite{Kat05}, in which the local spins
(magnetic moments) and the local polarization are coupled,
 \begin{eqnarray}
\vec{P} = \gamma \hat{e}_{ij} \times (\vec{\sigma}_i
\times \vec{\sigma}_j),
\label{polarization}
\end{eqnarray}
where $\hat{e}_{ij}$ is the unit vector connecting the neighboring
spins $\vec{\sigma}_i$ and $\vec{\sigma}_j$ with a material-dependent
coupling coefficient $\gamma$. Here we place the chain along the $x$
direction in the real space, i.e., $\hat{e}_{ij}=(1,0,0)$. Considering
a particular component ($z$ here, to be specific) of the spin
current,
\begin{eqnarray}
\frac{d \sigma_l^z}{dt}= i[{\cal H}, \sigma_l^z]=- {\rm div} j_l^z,
\end{eqnarray}
which defines the current $j_l^z$ and the corresponding $P_l^y$ by Eq.
(\ref{polarization}).

The {\it electric polarization} has two sources \cite{Men15}.
The first term originates from the spin-current model, given by
\begin{equation}
P_1^y \propto
\langle \sigma_l^x\sigma_{l+1}^y-\sigma_l^y\sigma_{l+1}^x\rangle,
\label{p1}
\end{equation}
which couples with $y$ component of the electric field $\vec{E}$
induced by the Dzyaloshinskii-Moriya interaction. Through the
relation $\vec{P_1}=(\partial{\cal H}/\partial\vec{E})$, the absence
of external electric field $\vec{E}$ in Hamiltonian $\cal{H}$ suggests
that it has little contribution to the electric polarization $P_1^y$.
However, as shown in Fig. \ref{OP_h}, $P_1^y$ is induced in the
presence of magnetic field $h$ as long as the phases are chiral,
and it is larger in chiral-II phase than in chiral-I phase.

\begin{figure}[t!]
\includegraphics[width=\columnwidth]{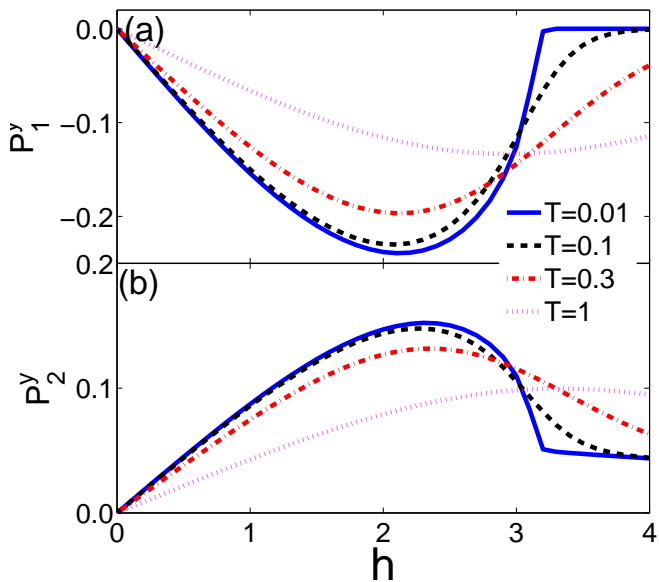}
\caption{
The evolution of electric polarization contributions $P_n^y$ with
increasing $h$ at different temperature $T$ for:
(a) $P_1^y$ (\ref{p1}), and
(b) $P_2^y$ (\ref{p2}).
Parameters are as follows: $J_o$=1, $J_e$=4, $\theta=\pi/3$, $J^*=0.5$.  }
\label{ChangeT}
\end{figure}

\begin{figure}[b!]
\includegraphics[width=\columnwidth]{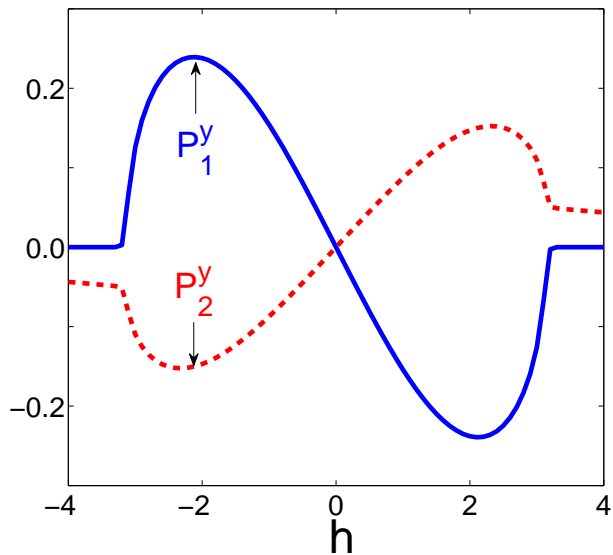}
\caption{The evolution of $P_1^y$ and $P_2^y$ by reversing the magnetic
field $h$. Parameters are as follows: $J_o$=1, $J_e$=4, $\theta=\pi/3$,
$J^*=0.5$, $T=0.01$.  }
\label{ReversedH}
\end{figure}

Another contribution of electric polarization may come from the spin
current triggered by the three-site interactions in the following way
\cite{Men15}:
\begin{equation}
P_2^y\propto - \langle \sigma_l^x\sigma_{l+1}^z\sigma_{l+2}^x
+\sigma_l^y\sigma_{l+1}^z \sigma_{l+1}^y\rangle.
\label{p2}
\end{equation}
The general form of the current operator is given in the Appendix.
The form of $P_2^y$ is the well-known XZX$+$YZY type of three-site
interaction and remains solvable in the frame of Jordan-Wigner
fermionization \cite{Tit03,Der11}. A little algebra will yield that
three-site XZX$+$YZY interactions acts here as a renormalization
(momentum-dependent) of the magnetic field $h$ in the Hamiltonian Eq.
(\ref{Mk_h}).

The manipulation of $h$ will affect finite $P_2^y$ in an indirect way,
as is displayed in Fig. \ref{OP_h}. We find that $P_2^y$ is also
induced by $h$, regardless of their phases. It has an opposite sign to
$P_1^y$ and almost complements its increase. Both $P_1^y$ and $P_2^y$
scale linearly with small $h$, indicating that they are triggered by
the external magnetic field. This is in contrast to some models with
two-spin interactions only, where the electric polarization can emerge
only for finite electric field. The compass model with three-site
interactions verifies the proposal in Ref. \cite{Bro13}, and indeed
exhibits a nontrivial magnetism-driven ferroelectricity.
We can observe in Fig. \ref{ChangeT} that the ferroelectricity
phenomena are quite stable for moderate temperature. An essential
feature of the ferroelectric behavior is that the electric polarization
can be reversed by the reversal of the magnetic field, as is verified
in Fig. \ref{ReversedH}.

\section{Summary and Conclusions}
\label{sec:summa}

In this paper we have considered the 1D generalized compass model
Eq. (\ref{Hamiltonian1}) which interpolates between the Ising model
(at $\theta=0$) and the maximally frustrated quantum compass model
(at $\theta=\pi/2$) and includes three-site XZY$-$YZX interactions.
We also investigated this model in the presence of external magnetic
field. Although the system is quantum and highly frustrated, we have
shown that exact solutions of the corresponding model may be obtained
through Jordan-Wigner transformation.

The XZY$-$YZX type of three-site interactions break both the parity
symmetry and the time-reversal symmetry, and then drastically modify
the energy spectra, leading to two kind of Tomonaga-Luttinger liquids.
We find that moderate three-site XZY$-$YZX interactions will lead to a
chiral-I state with two Fermi points in the representation of spinless
fermions, and large three-site XZY$-$YZX interactions transform the
system into the four Fermi point spinless fermions. Accordingly, this
modification of the Fermi surface topology follows some noticeable
changes in the central charges, and then affects the ground state
properties, such as nearest neighbor correlation functions. We find
that the $z$ component of scalar chirality operator can well distinguish
gapped and gapless phases, and also witness an abrupt change from
chiral-I to chiral-II phase. In both spin-liquid phases, not only the
magnetization is influenced by the magnetic field but the polarization
emerges even for $\vec{E}=0$ and is also affected by the magnetic field.

To conclude,
we emphasize that the advantage of the model considered here is its
exact solvability that implies in particular the possibility to
calculate accurately various dynamic quantities. The reported results
may serve to test other approximate techniques used to study more
realistic models.

\acknowledgments
W.-L.Y. acknowledges support by the Natural Science Foundation of
Jiangsu Province of China under Grant No. BK20141190 and
the NSFC under Grant No. 11474211.
A.M.O. kindly acknowledges support by Narodowe Centrum Nauki
(NCN, National Science Center) Project No. 2012/04/A/ST3/00331.

\appendix*

\section{Current operator for the compass model}

For a 1D compass chain, the only conserved quantity is the energy.
We can decompose Eq. (\ref{Hamiltonian1}) into:
\begin{equation}
H_{\rm GCM}(\theta)= \sum_{i=1}^{N'} h_i(\theta),
\label{Hamiltonian1-2}
\end{equation}
where
\begin{equation}
h_i(\theta)\!=
 J_{o}\tilde{\sigma}_{2i-1}(\theta)\tilde{\sigma}_{2i}(\theta)
+J_{e}\tilde{\sigma}_{2i}(-\theta)\tilde{\sigma}_{2i+1}(-\theta),
\label{hi}
\end{equation}
and $\tilde{\sigma}_i(\theta)$ in defined by Eq. (\ref{tilde}).
A unit cell contains
two bonds. Furtheron, one finds the commutation relations:
\begin{eqnarray}
&&[\tilde{\sigma}_{i}(\theta), \tilde{\sigma}_{j}(\theta)]=0,
\nonumber \\
&&[\tilde{\sigma}_{i}(\theta), \tilde{\sigma}_{j}(-\theta)]=
-2i\sin \theta  \sigma_i^z \delta_{ij},
\nonumber \\
&&[\tilde{\sigma}_{i}(-\theta), \tilde{\sigma}_{j}(\theta)]=
2i\sin \theta  \sigma_i^z \delta_{ij},
\nonumber \\
&&[\tilde{\sigma}_{i}(-\theta), \tilde{\sigma}_{j}(-\theta)]=0.
\end{eqnarray}

The energy current $\hat{J}_l$ of a compass chain in the nonequilibrium
steady states is calculated by taking a time derivative of the energy
density and follows from the continuity equation \cite{Zotos97,Antal97}:
\begin{eqnarray}
\frac{d h_l}{dt}&=& i[{\cal H}, h_l]\nonumber \\
&=& 2J_o J_e \sin\theta \left(
\tilde{\sigma}_{2l}(-\theta)\sigma_{2l+1}^z\tilde{\sigma}_{2l+2}(\theta)
\right.\nonumber \\
&-&\left.
\tilde{\sigma}_{2l-2}(-\theta)\sigma_{2l-1}^z\tilde{\sigma}_{2l}(\theta)
\right)\nonumber \\
&=&-(\hat{J}_{l+1}-\hat{J}_{l})=- {\rm div} \hat{J}_l,\\
\hat{J}_l&=&-2J_o J_e \sin\theta \tilde{\sigma}_{2l-2}
(-\theta)\sigma_{2l-1}^z\tilde{\sigma}_{2l}(\theta).
\end{eqnarray}
This energy current operator acts on three adjacent sites and has the
$z$ component of spin-1/2 operators between two odd sites. It depends
on $\theta$ in general. For $\theta$=0, it will present an
XZX type, while it exhibits a XZY type for $\theta=\pi/2$ \cite{Robin}.
For simplicity we choose $\theta=\pi/2$ in this term while still keep
$\theta$ as an arbitrary variable in the compass chain.

Of course the odd choice of operators is artificial and follows from
construction. If one defines,
\begin{equation}
h_l(\theta)=
 J_{e}\tilde{\sigma}_{2l}(-\theta)\tilde{\sigma}_{2l+1}(-\theta)
+J_{o}\tilde{\sigma}_{2l+1}(\theta)\tilde{\sigma}_{2l+2}(\theta),
\end{equation}
to replace Eq. (\ref{hi}), one finds
\begin{eqnarray}
\hat{J}_l=2J_o J_e \sin\theta \tilde{\sigma}_{2l-1}(\theta)
\sigma_{2l}^z\tilde{\sigma}_{2l+1}(-\theta).
\end{eqnarray}
The above dependence proves that the "macroscopic" current,
$\hat{J}=\sum_i \hat{J}_i$, will manifest
itself in the presence of an effective field $J^*$,
\begin{equation}
H= H_{\rm GCM}(\theta) -J^* \sum_i \hat{J}_i.
\label{Hamiltonian1-3}
\end{equation}

\end{document}